# Development and demonstration of next generation technology for Nb$_3$Sn accelerator magnets with lower cost, improved performance uniformity, and higher operating point in the 12-14 T range

Giorgio Ambrosio*, Giorgio Apollinari, Vito Lombardo, Stoyan Stoynev, Mauricio Suarez, George Velev
*Fermi National Accelerator laboratory, Batavia, IL 60510*
*Corresponding author: giorgioa@fnal.gov*

Paolo Ferracin, Soren Prestemon, GianLuca Sabbi
*Lawrence Berkeley National Laboratory, Berkeley, CA 94720*

Kathleen Amm
*Brookhaven National Laboratory, Upton, NY 11973*

**Executive summary**

The scope of the proposal outlined in this white paper is the development and demonstration of the technology needed for next generation of Nb$_3$Sn accelerator magnets in the 12-14 T range. The main goal is to cut magnet cold-mass cost by a factor 2 or higher with respect to the Nb$_3$Sn magnets produced by the US Accelerator Upgrade Project (AUP) [1] for the High-Luminosity Large Hadron Collider (HL-LHC) [2].  This goal will be achieved by significant reduction of labor hours, higher operating point, and improved performance uniformity. A key factor will be automation that will be achieved through industry involvement and benefitting from the experience gained in US national laboratories through the production of the AUP magnets.  This partnership will enable the development of a technology that will be easily transferable to industry for mid- and large-scale production of Nb$_3$Sn accelerator magnets in the 12-14 T range.  This step is essential to enable next generation of colliders such as the FNAL-proposed Muon Collider, FCC and other HEP hadron colliders.

This is a "Directed" R&D where direction is given by the field range and industry involvement for high-automation and industry-ready technology. The plan includes ten milestones, to be achieved in 6-8 years at the cost of 5-7 $M/year.

**Introduction**

Following the successful start of the LHC in 2010 and the Nobel-prize discovery of the Higgs boson in 2012, the LHC has continued to shed light on some of the fundamental physics questions of the age: the existence, or not, of supersymmetry; the nature of dark matter; the existence of extra dimensions; as well as the properties of the Higgs boson itself.

An improvement of the LHC and its detectors, called HL-LHC [2], has been approved in 2016 to allow the full exploitation of the LHC in the third and fourth decade of this century and to allow unique research opportunities both in fundamental discoveries and in accelerator science.

The United States government is making an investment of more than $750 M in the upgrade of the LHC to achieve the High Luminosities necessary to fully exploit the HEP frontier at the LHC energies. These investments will support the construction of upgraded detectors (CMS, ATLAS) and the construction of

new Interaction Regions (IR) designed to support a 10-fold increase of the luminosity delivered to the detectors. The upgraded machine, called HL-LHC, is presently in its construction phase. The US is contributing to the Accelerator part of HL-LHC through a DOE approved Project called Accelerator Upgrade Project (AUP) [1], executed "in-house" by a collaboration of National Labs (FNAL, BNL, LBNL) and to be deployed at CERN for installation and commissioning of HL-LHC in the 2026-2029 period.

It is a historical fact that the feasibility of the HL-LHC Project was made possible by the high-impact DOE investment in the LHC Accelerator R&D Program (LARP) [3-4], mandated by OHEP and executed in 2003-2016 as a directed R&D Program. This program aimed at developing appropriate technologies for what was seen, in the early 2000's, as an inevitable upgrade of the LHC capabilities.

This 2022 Snowmass Process plans to map the road for High Energy Physics in the upcoming decades for the US HEP community on a global scale. Among the elements of possible future plans, advanced Colliders (Muon Collider [5-6], HE-LHC, next-generation Hadron Colliders such as FCC-hh [7-8], etc.) will play central roles in the exploration of the Energy Frontier. The circular nature of the Colliders under consideration naturally drives the focus to high field magnet technology to be used in the main ring arcs.

A critical element of the technology covered by this White Paper is the challenge of lower costs, higher conductor efficiency (i.e. less conductor for same field in magnet aperture), performance uniformity at large scale, and higher production efficiencies than those presently achieved in the "boutique-like enterprise" of HL-LHC [9-11].

We therefore envision and propose a "Directed" R&D program, to be executed in the next 6-8 years, that would develop and demonstrate next generation technology for $Nb_3Sn$ accelerator magnets with lower cost, better performance reproducibility and higher efficiency than presently achievable in the 12-14 T range. The overall goal is to cut costs by a factor of two or more with respect to the cost per meter of AUP magnets. A key factor will be automation that will be achieved through industry involvement from the very beginning. Industry involvement will make possible the development of a technology ready for smooth transfer to industry and will enable the construction of magnets to be produced in the 100's or 1000's for future advanced colliders, such as FNAL-proposed Muon Collider and other HEP hadron colliders.

We think that significant savings are possible based on the experience gained through the production of the AUP (MQXFA) magnets [10] for HL-LHC and considering that the present technology for $Nb_3Sn$ accelerator magnets was developed aiming at a single goal: achieving the required performance. This goal was justified for a limited number of accelerator magnets (MQXFA/B quadrupole and 11T dipole magnets) using a conductor that was introducing new challenges: the high strain sensitivity and the need of a high-temperature heat treatment. Nonetheless, at this time it is possible and necessary to use the experience gained in the last 20 years to rethink the whole technology used for the design and fabrication of $Nb_3Sn$ accelerator magnets and start the development of next generation technology.

The main goal of next generation technology is a significant reduction (by a factor of two or higher) of the overall magnet cost. This goal is achievable through reduction of touch labor in all fabrication steps, minimizing the required magnet operating margin (i.e. higher operating point on magnet load line resulting in higher field for less conductor), and higher coil/magnet yield. Feedback from industry and possible integration of industry components will be pursued during all phases of the proposed development.

The ultimate goal of improved and cost-efficient magnet fabrication is achievable by a US team (with industry involvement) because of the large production experience fabricating LARP and AUP (MQXFA) magnets. Simultaneously, the US industrial complex is going to gain a foothold in the acquired knowhow for accelerator magnet fabrication. MQXFA magnets, which have excellent documentation coverage, will be used as the benchmark to assess improvements. The plan presented in this paper includes phases and steps with milestones to demonstrate progress.

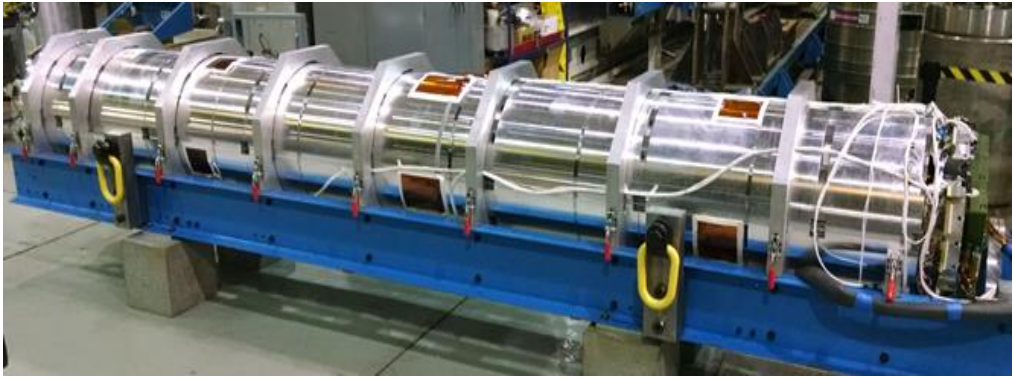

*Figure 1: MQXFA magnet on shipping frame*

*Table 1: Main parameters of MQXFA magnet*

| PARAMETER | Unit | MQXFA |
|---|---|---|
| COIL APERTURE | mm | 150 |
| MAGNETIC LENGTH | m | 4.2 |
| N. OF LAYERS | | 2 |
| N. OF TURNS INNER-OUTER LAYER | | 22-28 |
| OPERATION TEMPERATURE | K | 1.9 |
| NOMINAL GRADIENT | T/m | 132.2 |
| NOMINAL CURRENT | kA | 16.23 |
| PEAK FIELD AT NOM. CURRENT | T | 11.3 |
| SHORT SAMPLE LIMIT | | 76% |
| STORED ENERGY AT NOM. CURR. | MJ/m | 1.15 |
| DIFF. INDUCTANCE | mH/m | 8.26 |
| STRAND DIAMETER | mm | 0.85 |
| STRAND NUMBER | | 40 |
| CABLE WIDTH | mm | 18.15 |
| CABLE MID THICKNESS | mm | 1.525 |
| KEYSTONE ANGLE | | 0.4 |

**Cost analysis and overall plan**

Understanding cost drivers is critical for an effective plan to reduce costs. In the following we present the cost analysis of the $Nb_3Sn$ low-beta quadrupoles produced in the U.S. for the HL-LHC. This cost analysis is limited to the magnetic element, the "MQXFA magnet" (Fig. 1) without taking into consideration cold mass and cryostat, which are common to every accelerator magnet. The main MQXFA magnet parameters are reported in Table 1.

The data used for this analysis were taken from the cost-per-part assessment made by the US AUP project and used to estimate the cost of project risks. This data set presents a realistic, all-factors-included picture of the present cost of high-field $Nb_3Sn$ accelerator magnets from a US perspective (i.e., including personnel, yield assumptions, and some risk mitigation factors).

The results of the cost analysis are presented in Table 2. The numbers in bold represent subtotals per magnet, whereas the non-bolded numbers represent the cost per single cable or coil, the cost of a magnet vertical test and the cost of yield assumptions. The cost per cable (item 1) includes strand procurement, cable fabrication, cable insulation and QC.  The cost of coil fabrication (item 3) shows the additional costs (i.e. not included in item 1) to obtain a completed coil: winding, binder curing, heat treatment, epoxy impregnation, and all M&S for these fabrication steps. Item 5 includes structure procurement, structure subassemblies, magnet assembly and pre-load leading up to the complete magnet and overall QC.  Item 6 shows the cost estimate for a production vertical test assuming 20 training quenches at 1.9 K, magnetic measurements, two thermal cycles, and a few quenches for temperature and ramp-rate dependence studies.  Item 7 was obtained by computing the cost of all the production yield assumptions (additional cables, coils, magnet assemblies, and vertical tests) included in the project baseline and dividing it by the number of deliverable magnets. Items 6 and 7 may be grouped together because they are different strategies for addressing the same issue: some components (cables/coils/magnets) may fail and need replacement. The total cost of this group is about $1.1M and we identify it as "Yield cost".

Table 2 shows that the total cost of an MQXFA magnet, including yield cost is about $4.7 M. Considering that MQXFA magnetic length is 4.2 m, the resulting magnet cost per length is about 1.1 $M/m in this length range (4 - 5 m). Longer magnets allow for some cost savings because the labor needed for some activities does not scale with length.  Nonetheless, these savings are limited with the present technology and will be more significant with the improvements proposed by this program.
The total cost can be divided into four parts, each costing about $1.1-$1.2 M: 1) four insulated cables; 2) fabrication of four coils; 3) structure procurement and magnet assembly; and 4) the yield cost. M&S cost is the dominant component of Item 2 (i.e., conductor procurement) and Item 5 (i.e. structure parts procurement). Labor cost is the dominant component of Item 4 (i.e. coil fabrication labor time).

*Table 2: Cost analysis of MQXFA magnets by US LHC Accelerator Upgrade Project*

| Item | Part | Labor (k$) | M&S (k$) | Total (k$) |
|---|---|---|---|---|
| 1 | One insulated cable | 67 | 215 | 282 |
| 2 | **Four insulated cables** | **267** | **861** | **1,128** |
| 3 | One coil fabrication (w/o cable cost) | 241 | 60 | 301 |
| 4 | **Four coils fabrication (w/o cables cost)** | **963** | **240** | **1,203** |
| 5 | **Structure procurement & Magnet assembly** | **276** | **973** | **1,249** |
| 6 | Magnet Vertical test | | | 603 |
| 7 | Yield assumptions (cost per magnet) | | | 519 |
| 8 | **Yield cost** | | | **1,122** |
| | **TOTAL per MQXFA magnet** | | | **4,703** |

The MQXFA cost analysis shows that there is not a single main cost driver for these magnets, but it also shows that labor costs make ~1/3 of the total magnet cost and ~42% of magnet fabrication itself. Even if different $Nb_3Sn$ accelerator magnets may have other cost distributions, the big picture is not expected to change significantly. This observation may be employed to reach the following conclusion*: In order to achieve significant cost reductions, the next generation of $Nb_3Sn$ accelerator magnets must be developed with the goal of cost reduction in the following areas: 1) conductor cost, 2) labor for coil fabrication, 3) structure components cost, and 4) yield cos*t.

The plan presented in this white paper aims at addressing all cost drivers, starting from coil fabrication technology up to yield costs. Some goals will be achieved through the effort planned in this proposal, whereas other goals will be achieved through synergies with other R&D programs (such as the US Magnet Development Program [12], Early Carrier Awards, Laboratory Directed R&D efforts) and demonstrated through the program proposed in this white paper.  Automation will be key factor, and industry will be involved from the very beginning to develop solutions compatible with effective industrial production. The largest cost reductions (more than 50% per area) are expected in coil fabrication (because of automation) and yield cost (because less touch labor will reduce risk of component failures). Significant cost reduction is expected also in structure procurement and magnet assembly because of feedback from industry expertise. Therefore, the overall cost reduction is expected to exceed 50%.

The proposed program is made of three main phases described below.  Progress will be measured against ten milestones. Phases 1 and 2 are expected to take two or three years depending on the level of funds and available resources.  Phase 3 has five milestones and it will take the whole duration of this program that is expected to be 6-8 years. Program cost is estimated at $5-7 M per year based on similar work by LARP and on extrapolations for industry involvement.

This program is self-standing, nonetheless it may also be included in the *Leading-Edge technology And Feasibility-directed (LEAF) Program aimed at readiness demonstration for Energy Frontier Circular Colliders (pp, $\mu\mu$) by the next decade* [13].

**Phases and Milestones**

**Phase1**:
The goal of phase 1 is the development and demonstration of MQXF-like coils, which can be fabricated requiring ≤ 50% touch labor than present MQXFA coils. Significant cost reduction will be achieved by lower labor time per coil and by higher coil yield. The latter will be the outcome of reduced risk of coil damage during fabrication.
In this phase we will use mostly AUP leftover conductor, coil fabrication tooling (although with some modifications), and magnet structures. This phase will benefit from conditions not easily found otherwise: assembly teams with solid know-how, available materials, tooling and infrastructure.
Industry will be involved in this phase from the very beginning focusing on automation and processes well fit for industrial production.  Nonetheless, the goal of this phase is not just industrialization of the coil fabrication technology presently used for $Nb_3Sn$ accelerator magnets (e.g. MQXF or 11T magnets).  On the contrary, we are going to develop a new coil fabrication technology requiring significantly less touch labor and optimized for transfer to industry. Industry input is needed for both goals and will be a key component of this plan. The actual transfer to industry should happen later through a dedicated program or a construction project.

In this phase we will also understand what the most productive way is to engage industry for all three phases.

Several directions will be explored during the initial development including feedback from other R&D programs. Examples are: coil winding and handling without use of ceramic binder, fabrication of single layer coils with layer-layer splice in order to reduce by half cable/coil loss in case of issue during production, highly automated coil fabrication reducing tech hours and increasing the production yield, reduction of coil handling from heat treatment to epoxy impregnation, use of tougher and high-Cp (Heat capacity) impregnation materials reducing magnet training, cable cleaning for good bonding to impregnation material, and tooling/process optimization for minimizing damage risks based on analysis of MQXFA coil production.

**Phase 1 milestones are:**
> Milestone 1: fabrications of five MQXF-like short coils with less than 50% tech hours used for the fabrication of MQXF coils.
> Milestone 2: test of these coils in MQXF mirror structure and short-model structure, and demonstration of performance (training, $I_{max}$/ $I_{SSL}$, coil reproducibility) similar or better than MQXF short models.

**Phase 2:**
The main goal of Phase 2 is the demonstration that enhanced conductor can be used with the technology developed in phase 1. The enhanced conductor is expected to be handed off by other programs and must have higher heat capacity and possibly higher critical current than the conductor used by AUP in MQXFA coils. Candidate for such conductor is currently developed $Nb_3Sn$ wire with Artificial Pinning Centers (~ 50% increasing of Jc comparing to AUP conductor) [14] and high-Cp additives (increasing by several times the wire minimum quench energy) [15]. This phase will use coil fabrication technology and tooling developed during Phase 1, together with conductor developed by other programs and institutes (MDP, CPRD, early career awards, NHMFL), for fabrication of MQXF-like coils with a new enhanced conductor. The enhanced conductor with higher heat capacity is expected to allow shorter training and higher operating point than MQXF short models assembled with the same pre-stress level. This phase is an opportunity for synergies with other programs, which will benefit from testing the enhanced conductor against a well-established benchmark.

**Phase 2 milestones are:**
> Milestone 3: fabrications of five MQXF-like short coils with less than 50% MQXF-coil tech hours, using enhanced conductor.
> Milestone 4: test of these coils in MQXF mirror structure and short-model structure, and demonstration of enhanced performance: $I_{max}$ > 85% $I_{SSL}$ (i.e. > 10% better than MQXFA magnets).

**Phase 3:**
The main goal of Phase 3 is the development and demonstration of a high efficiency, low cost $Nb_3Sn$ accelerator magnet in the 12-14 T range. This phase will use coil fabrication technology and conductor demonstrated in Phase 1 and 2, together with a low-cost magnet structure. Actual conductor dimensions and coil fabrication tooling may need to be adjusted for the phase-3 magnet design. The overall magnet design will depend on the P5 recommendations and may be developed in collaboration with other R&D programs.

The design and development of this magnet should be done in parallel with Phases 1 and 2 work, in order to be ready for coil fabrication and structure procurement as soon as Phase 2 work is complete. In this phase industry will be involved aiming at reducing cost of structure components, development of magnet assembly procedures with limited touch labor and easily transferable to industry.

The actual technology transfer should be completed by a dedicated program or project at later time. This phase aims at development of all elements needed for low-cost $Nb_3Sn$ accelerator magnets in the 12-14 T range, and demonstration of feasibility. In case P5 recommendations stress the need for large aperture magnets (for instance for a Muon Collider), the phase-3 magnet will have a large aperture and possibly slightly lower magnetic field in order to keep coil stresses within acceptable limits.

**Phase 3 milestones are:**
> Milestone 5: Design of a low-cost $Nb_3Sn$ accelerator magnet according to P5 recommendations.
> Milestone 6: Fabrication of a set of coils with standard conductor and technology developed in phase 1 for use in low-cost $Nb_3Sn$ magnet.
> Milestone 7: Engineering design and procurement of low-cost magnet structure.
> Milestone 8: Assembly and test of low-cost $Nb_3Sn$ accelerator magnet with coils from Milestone 6 and structure from Milestone 7.
> Milestone 9: fabrication of 2 sets of coils with enhanced conductor demonstrated in phase 2 for use in low-cost $Nb_3Sn$ magnet.
> Milestone 10: Assembly and test of two low-cost high-efficiency $Nb_3Sn$ accelerator magnets with coils from Milestone 9.

It should be noted that to achieve these milestones more test coils and model magnets will have to be fabricated and tested than those listed in the milestones. The total number will be optimized later and may depend on schedule constrains and available funding. For instance, tighter schedule will require parallel developments and overall larger number of coils and magnet tests. Nevertheless, this plan should be adequate to demonstrate program goals and draft an optimal path for development and collaboration with industry.

**Conclusion**

The proposed program will enable development and demonstration of next generation $Nb_3Sn$ accelerator magnets with lower cost, improved performance uniformity, and higher operating point than present $Nb_3Sn$ magnets in the 12-14 T range (e.g. MQXF and 11T magnets). It will include automation and industry involvement as critical elements together with the expertise and know-how of the US labs, which are producing $Nb_3Sn$ magnets for the AUP project. The next-generation technology will have significantly lower cost than the present one, by a factor two or more, and is needed for mid and large-scale production of $Nb_3Sn$ accelerator magnets (for instance for FNAL-proposed Muon Collider and other HEP hadron colliders). The materials, technologies, assembly solutions, know-how and expertise developed for this program will leverage other programs and projects, such as the development of higher field (15-16 T) $Nb_3Sn$ accelerator magnets and hybrid magnets (~20 T).

While we strongly believe that the path we suggest, building upon the US AUP project, is optimal for significant cost-reduction in production of $Nb_3Sn$ accelerator magnets, we emphasize that a directed program of this kind is necessary for High Energy Physics.